\documentclass[a4paper,preprint,12pt]{revtex4}
\usepackage{epsfig}
\usepackage{amssymb,amsmath}
\usepackage{mathrsfs}
\usepackage{bbm}
\usepackage{graphicx,subfigure}

\newcommand{\dsl}{\displaystyle{\not}}

\begin{document}

\title{\bf   $\bar{B} \to X_s \gamma$ constraints on the top quark anomalous $t\to c\gamma$ coupling }
\author{Xingbo Yuan$^{1}$, Yang Hao$^{1}$  and Ya-Dong Yang$^{1,2}$\\
{ $^1$\small Institute of Particle Physics, Huazhong Normal University, Wuhan, Hubei  430079, P.~R. China}\\
{ $^2$\small Key Laboratory of Quark \& Lepton Physics, Ministry of Education, Huazhong Normal University,}\\
{\small Wuhan, Hubei, 430079, P.~R. China}}

\bigskip\bigskip
\vspace{-1.5cm}

\begin{abstract}
Observation of  top quark flavor changing neutral process $t\to c +\gamma$ at the LHC would be the signal of physics beyond the Standard Model. If  anomalous $t\to c\gamma$ coupling  exists, it will affect the precisely measured $\mathcal{B}(\bar{B} \to X_s \gamma)$.  In this paper, we study the effects of a dimension 5 anomalous $ tc \gamma$  operator  in  $\bar{B} \to X_s \gamma$ decay to derive constraints on its possible strength.  It is found that, for real  anomalous $t\to c\gamma$ coupling $\kappa_{\rm{tcR}}^\gamma$, the constraints correspond to the upper bounds $\mathcal B ( t \to c + \gamma )<6.54 \times 10^{-5}$ (for $\kappa_{\rm{tcR}}^\gamma>0$) and  $\mathcal B ( t \to c + \gamma )<8.52 \times 10^{-5}$ (for $\kappa_{\rm{tcR}}^\gamma<0$), respectively, which  are about the same order as the $5\sigma$ discovery potential  of ATLAS ($9.4\times 10^{-5}$) and slightly lower than that of CMS ($4.1\times 10^{-4}$) with $10 \ \rm{fb}^{-1}$ integrated luminosity operating at $\sqrt{s} =14$ TeV.
\end{abstract}

\maketitle
\newpage

\section{Introduction}
\label{sec:intro}

In the Standard Model (SM), top quark lifetime is dominated by the $t\to b W^+$ process, and its flavor changing neutral current (FCNC) processes $t \to q V (q=u,c; V=\gamma, Z, g)$ are extremely suppressed by GIM mechanism.  It is known that the SM predicts very tiny top FCNC branching ratio  $\mathcal B(t \to q V)$,  less than $\mathcal{O}(10^{-10})$ \cite{gadi}, which would be  inaccessible at the CERN Large Hadron Collider(LHC). In the literature \cite{gadicp,Beneke}, however, a number of interesting questions have been intrigued by the large top quark mass which is close to the scale of electroweak symmetry breaking. For example, one may raise the question whether new physics (NP) beyond the SM could manifest itself in nonstandard couplings of top quark which would show up as anomalies in the top quark productions and decays.

At present, the direct constraints on $\mathcal B(t \to q V)$ are still very weak.  For its radiative decay, the available experimental  bounds are $\mathcal B(t \to u \gamma)<0.75\%$ from ZEUS \cite{ZEUS} and
$\mathcal B(t \to q \gamma)<3.2\%$ from CDF \cite{CDF} at $95\%$ C.L., respectively.  These constraints will be improved greatly by the large top quark sample to be available at the LHC, which is expected to produce  $8\times 10^6$  top quark pairs  and another few million single top quarks  per year at low luminosity ($10 \ \rm{fb}^{-1}$/year). Both ATLAS
\cite{ATLAS} and CMS \cite{CMS} have got analyses ready for hunting out top quark FCNC processes as powerful  probes for NP.  With $10 \ \rm{fb}^{-1}$ data, it is expected that  both ATLAS and CMS  could observe  $t\to q \gamma$  decays if their branching ratios are enhanced to $\mathcal {O}(10^{-4})$ by anomalous top quark couplings \cite{ATLAS, CMS}. However, if the top quark anomalous couplings present,  they will affect some precisely measured qualities  with virtual top quark contribution. Inversely, these qualities can also restrict the possible number of  top quark FCNC decay signals at the LHC. The precisely measured inclusive decay $B \to X_s \gamma$ is one of the well known sensitive probes for extensions of the SM, especially the NPs which alter the strength of FCNCs \cite{top}. Thus, when performing the study of the possible strength of $t\to c\gamma$ decays at the LHC, one should take into account the constraints from  $B \to X_s \gamma$ \cite{wtb,Fox}.

In this paper, we will study the contribution of anomalous $t\gamma c$ operators  to the $\bar{B} \to X_s\gamma$
 branching ratio and derive  constraints on its strength.  In the next section, after a brief discussion of  a set of  model-independent dimension 5 effective operators relevant to $t\to c \gamma$ decay, we calculate the effects of operator  $ \bar c_L \sigma^{\mu \nu} t_R F_{\mu \nu}$  in $B \to X_s \gamma$ decay, which result in a modification to $C_{7\gamma}$.  In Sec. III we present  our numerical results of  the constraints on its strength  and  the  corresponding upper limits on branching ratio of $t\to c \gamma$ decays. Finally,  conclusions are made in Sec. IV.  Calculation details are presented in Appendix A, and input parameters are collected in Appendix B.

\section{Top quark anomalous couplings  and their effects in $\bar{B} \to X_s\gamma$ decay}

Without resorting to the detailed flavor structure of a specific  NP model,  the Lagrangian describing the top quark anomalous couplings can be written in a model independent way with dimension 5 operators \cite{lag1}
\begin{align}
{\mathcal L}_5 =
&-  g_s \sum_{q=u,c,t}
\frac{\kappa ^g_{tqL}}{\Lambda} \bar q_R \sigma^{\mu \nu}T^a t_L G^a_{\mu \nu}
-  \frac{g}{\sqrt 2}  \sum _{q=d,s,b}
\frac {\kappa ^W_{tqL}}{\Lambda} \bar q_R \sigma^{\mu \nu} t_L W^-_{\mu \nu}
-e \sum_{q=u,c,t}  \frac{\kappa ^\gamma _{tqL}}{\Lambda} \bar q_R \sigma^{\mu \nu} t_L F_{\mu \nu}
 \nonumber\\
&- \frac {g}{2\cos \theta _W}  \sum _{q=u,c,t} \frac {\kappa ^Z_{tqL}}{\Lambda} \bar q_R \sigma^{\mu \nu} t_L Z_{\mu \nu}
+(R \leftrightarrow L)+h.c.,
\label{lag1}
\end{align}
where $\kappa$ is the complex coupling of its corresponding operator, $\theta _W$ is the weak angle, and $T^a$ is the Gell-Mann matrix. $\Lambda$ is the possible new physics scale, which is unknown but may be much larger than the electroweak scale. There are also Lagrangian describing the top quark anomalous interactions with dimension 4 and 6 operators, and the dimension 4 and 5 terms can be traced back to dimension 6 operators \cite{wyler,list}. In fact top quark anomalous interactions can be generally described by the gauge-invariant effective Lagrangian with dimension 6 operators in a form without redundant operators and parameters \cite{Fox,Saavedra}. A recent full list of dimension 6 operators  could be found in Ref.~\cite{SM6}. But for on-shell gauge bosons, the Lagrangian in Eq.~(\ref{lag1}) works and is commonly employed in high energy phenomenology analysis \cite{Beneke, ATLAS,Li}.

The operators in Eq.~(\ref{lag1}) relevant to  $t\to q \gamma $ decays  read
\begin{eqnarray}
{\mathcal L}_\gamma  =
-  e  \sum _{q=u,c}
 \frac {\kappa ^\gamma _{tqL}}{\Lambda} \bar q_R \sigma^{\mu \nu} t_L F_{\mu \nu}-  e  \sum _{q=u,c}
 \frac {\kappa ^\gamma _{tqR}}{\Lambda} \bar q_L \sigma^{\mu \nu} t_R F_{\mu \nu}+h.c..
 \label{lag2}
\end{eqnarray}
 It is understood that the Dirac matrix $\sigma_{\mu\nu}$ connects left-handed fields to right-handed
 fields, the $t\to c \gamma$ transition will involve two independent operators
  $m_{q}\bar q_R \sigma^{\mu \nu} t_L F_{\mu \nu}$  and $m_{t}\bar q_L \sigma^{\mu \nu} t_R F_{\mu \nu}$,
  where the mass factors must appear whenever a chirality flip $L\to R$ or
 $R\to L$ occurs. Due to the mass hierarchy $m_{t}\gg m_{c}$, the effect of $m_{q}\bar q_R \sigma^{\mu \nu} t_L F_{\mu \nu}$  can be neglected unless $\kappa ^\gamma _{tqL}$ is enhanced to be comparable to $\tfrac{m_{t}}{m_{c}}\kappa ^\gamma _{tqR}$ by  unknown mechanism.

\begin{figure}[t]
\centering
\subfigure[]{\includegraphics [width=5cm]{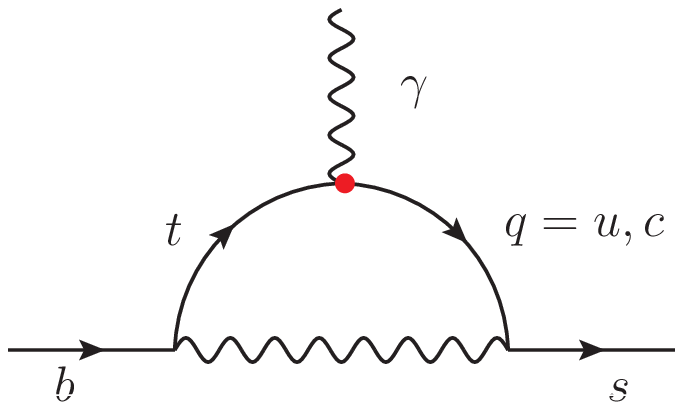}\label{fig1a}}
\subfigure[]{\includegraphics [width=5cm]{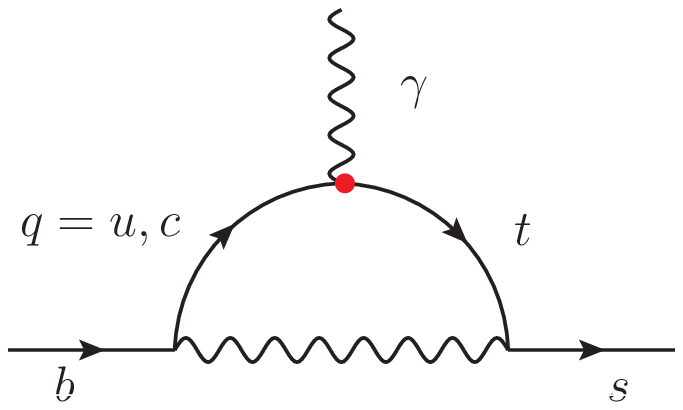}\label{fig1b}}
\subfigure[]{\includegraphics [width=5cm]{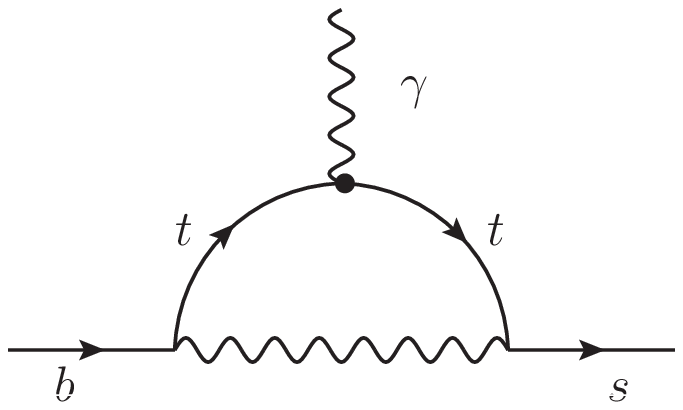}\label{fig1c}}
\caption{ Feynman diagrams for  $b \to s \gamma$.  (a) and (b) are the penguin diagrams with the anomalous 
 $t q \gamma $ couplings. (c) Sample LO penguin diagram in the SM.}\label{fig1}
\end{figure}

The anomalous $t \gamma q$ coupling  affects  $b \to s \gamma$ decays through the two Feynman diagrams depicted in  Figs.~\ref{fig1a} and \ref{fig1b}.  It is interesting to note that the CKM factors in  Fig.~\ref{fig1a} and Fig.~\ref{fig1b} are $V_{tb} V_{qs}^{*}$ and $V_{qb} V_{ts}^{*}$,  respectively.  Since $|V_{tb} V_{qs}^{*}| \gg  |V_{qb} V_{ts}^{*}|$ for $q=u,c$, the contribution of Fig.~\ref{fig1a} would be much stronger than that of Fig.~\ref{fig1b}.  Furthermore, given the strengths of $t\to u\gamma$ and  $t\to c\gamma$ comparable,  the contribution of Fig.~\ref{fig1a} to $b\to s \gamma$ is still dominated by $t\to c\gamma$  because of $|V_{cs}|\gg |V_{us}|$. Hence we will only consider  Fig.~\ref{fig1a} with anomalous
$t c \gamma$ coupling.
 From the Feynman diagram of  Fig.~\ref{fig1a}, it is easy to observe  that the large CKM factor $V_{tb}V_{cs}\approx 1$ makes  $b\to s \gamma$ very sensitive to the strength of  anomalous $t c \gamma$ coupling.

The calculation of Fig.~\ref{fig1a} can be carried out straightforwardly. The calculation details  are presented in Appendix A, and  the final  result reads
\begin{eqnarray}
i \mathcal{M}(b \to s \gamma) &=& \bar s [ e \Gamma ^{\nu} (k) ] b  \epsilon _{\nu}(k), \nonumber \\
e \Gamma^{\nu}(p,k)  &=& i e \frac {G_F}{4\sqrt 2 \pi^2} V^{*}_{cs} V_{tb}
\left[ i \sigma ^{\nu \mu} k_{\mu} (m_s f_{L}(x) L + m_b f_{R}(x) R)
\right].
\label{main}
\end{eqnarray}
Usually $m_{s}$ term can be neglected,  and the function $f_{\rm{R}}(x)$ is calculated to be
\begin{equation}
 f_{\rm {R}}(x) =  \frac{\kappa ^\gamma _{\rm{tcR}}}{\Lambda} 2m_{t}
 \left[-\frac{1}{(x_c-1)(x_t-1)}
 - \frac{x_c^2}{(x_c-1)^2(x_c-x_t)} \ln x_c + \frac{x_t ^2}{(x_t -1)^2 (x_c -x_t)} \ln x_{t}
 \right],
\end{equation}
 with $x_{q}=m_{q}^2/m_{W}^2$. Now we are ready to incorporate the NP contribution into its SM counterpart for ${\bar B}\to X_{s}\gamma$ decay.

In the SM, it is known that  ${\bar B}\to X_{s}\gamma$ decay is governed by the effective Hamiltonian  at scale
$\mu=\mathcal{O}(m_{b})$  \cite{Buras1}
\begin{eqnarray}
\mathcal{H}_{\rm{eff}}(b\to s \gamma)=-\frac{4G_{F}}{\sqrt 2}V_{ts}^* V_{tb}
\left[
  \sum_{i=1}^{6} C_{i}(\mu) Q_{i}(\mu)+C_{7\gamma} (\mu)O_{7\gamma}(\mu)+  C_{8g}(\mu) O_{8g}(\mu)
  \right],
  \label{basis}
\end{eqnarray}
where  $C_{i}(\mu)$ are the  Wilsion coefficients,  $O_{i=1-6}$ are the effective four quark operators and
\begin{eqnarray}
O_{7\gamma}=\frac{e}{16\pi^2} m_b (\bar s_L \sigma^{\mu \nu} b_R) F_{\mu \nu}, ~~~~
O_{8g}=\frac{g}{16\pi^2} m_b (\bar s_L \sigma^{\mu \nu} T^a b_R)G _{\mu \nu} ^{a}.
 \label{O}
\end{eqnarray}
For calculating $\mathcal{B}(\bar B \to X_{s}\gamma)$, instead of the original Wision coefficients $C_{i}$, it is convenient to
use the so called ``effective coefficients'' \cite{Buras94}
\begin{eqnarray}
C_{7\gamma}^{(0) \rm{eff}}(m_b)=\eta^{\frac{16}{23}}  C_{7 \gamma} ^{(0)\rm{SM}}(M_W)
+ \frac{8}{3}(\eta^{\frac{14}{23}} - \eta ^{\frac{16}{23}}) C_{8g}^{(0)\rm{SM}}(M_W)+ C_{2} ^{(0)\rm{SM}}(M_W)\sum_{i=1} ^8 h_i \eta^{a_i},
\label{C7eff}
\end{eqnarray}
where $\eta = \alpha_s(\mu_W)/\alpha_s(\mu_b)$ and
\begin{align}
h_i&=\bigl(&\tfrac{626126}{272277}&\ &  -\tfrac{56281}{51730}&\ &  -\tfrac{3}{7}&\ & & -\tfrac{1}{14} & -&0.6494 & & -0.0380 & & -0.0185 &  -0.0057&\ &\bigr),\\
a_i&=\bigl(&\tfrac{14}{23}&\ & \tfrac{16}{23}&\ & \tfrac{6}{23} &\ & & -\tfrac{12}{23} & & 0.4086 & &-0.4230 & & -0.8994 &  0.1456&\ &\bigr).
\end{align}
To the leading order approximation, the $\mathcal B (\bar B \to X_{s} \gamma)$ is proportional to
 $|C_{7\gamma}^{(0)\rm{eff}}(m_{b})|^2$ \cite{Buras}.

In terms of the operator basis in Eq.~(\ref{basis}), the contribution of the anomalous $t\to c\gamma$ couplings
 in Eq.~(\ref{main}) would result in  the deviation of
\begin{equation}
C_{7\gamma}(M_{W}) \to  C^{\prime}_{7\gamma}(M_{W})= C_{7\gamma}(M_{W})+C^{\rm{NP}}_{7\gamma}(M_{W})
\end{equation}
 and $C^{\rm{NP}}_{7\gamma}(M_{W})$ can
be read from Eq.~(\ref{main}) as
\begin{equation}
C_{7 \gamma} ^{\rm{NP}}(M_{W}) =
\frac{\kappa _{\rm{tcR}} ^{\gamma}}{\Lambda} \frac{V _{cs} ^*}{V_{ts} ^*}
m_{t}
\left[
\frac{1}{(x_c-1)(x_t-1)}
 + \frac{x_c^2}{(x_c-1)^2(x_c-x_t)} \log x_{c}-\frac{x_t ^2}{(x_t -1)^2 (x_c -x_t)} \ln x_{t}
 \right].
 \end{equation}
From this equation, one can see that the NP contribution is suppressed by a factor of $m_{t}/\Lambda$  but enhanced by $V_{cs}/V_{ts}$.

Since NP contribution does not bring about any new operator, the renormalization group  evolution of $C_{7\gamma}^{\rm eff}$ from  $M_{W}$ to $m_{b}$ scale is just the  same as the SM one in Eq.~(\ref{C7eff}).  For $m_{t}=172$ GeV, $m_{b}=4.67$ GeV, $\alpha_{s}(M_{Z})=0.118$ and $\Lambda=1$ TeV, we have
\begin{eqnarray}
C_{7\gamma}^{\prime \rm{eff}}(m_{b})&=&\eta^{\frac{16}{23}}
\left[ C_{7 \gamma} ^{(0)\rm{SM}}(M_W) + C_{7 \gamma} ^{(0)\rm{NP}}(M_{W})
\right]
+ \frac{8}{3}(\eta^{\frac{14}{23}} -
 \eta ^{\frac{16}{23}}) C_{8g}^{(0)\rm{SM}}(M_W)+ C_{2} ^{(0)\rm{SM}}(M_W)\sum_{i=1} ^8 h_i \eta^{a_i}
\nonumber\\
&=&0.665 \left[  C_{7 \gamma} ^{(0)\rm{SM}}(M_W) + C_{7 \gamma} ^{(0)\rm{NP}}(M_{W})
		\right]
+ 0.093 \ C_{8g}^{(0)\rm{SM}}(M_{W})-0.158 \ C_{2} ^{(0)\rm{SM}}(M_{W})
\nonumber\\
&=&0.665 \left[-0.189 + \kappa_{\rm{tcR}}^\gamma(-1.092) \right]
+ 0.093 \ (-0.095)-0.158.
\label{c7np}
\end{eqnarray}

In principle,  $C_{7\gamma}^{\prime \rm{eff}}(m_b)$ will receive corrections from anomalous $t\to c g$  couplings
 in Eq.~(\ref{lag1}) which will cause a deviation to $C_{8g}^{(0)\rm{SM}}(M_{W})$. However, as shown by Eq.~(\ref{c7np}), the  coefficient $\eta^{\frac{16}{23}}$ of  $C_{7 \gamma} ^{(0)}(M_{W})$ is about one order larger than  $
\frac{8}{3}(\eta^{\frac{14}{23}} - \eta ^{\frac{16}{23}})$ of $C_{8g}^{(0)\rm{NP}}(M_W)$.
Given the relative strength of  $C_{8g}^{(0)\rm{NP}}(M_{W})$ to $C_{8g}^{(0)\rm{SM}}(M_{W})$
at $10\%$ level,  $C_{7\gamma}^{\prime \rm{eff}}(m_b)$ will be shifted by only few percentage. For simplifying the numerical analysis, we would neglect the contribution of the anomalous $t\to c g$  couplings.   We  also find  that the operator
$\bar q_R \sigma^{\mu \nu} t_L F_{\mu \nu}$ contributes to $\bar B \to X_s\gamma $
only through the term $m_{s} \bar s \sigma_{\mu\nu}  (1-\gamma_{5})b$ as shown by Eq.~(\ref{main}) and Eq.~(\ref{a7}).
Combined with the previous remarks on this operator,   the effects of  $\bar q_R \sigma^{\mu \nu} t_L F_{\mu \nu}$
could be safely neglected.

\section{Numerical results and discussions}
The current average of experimental results of   $\mathcal B (\bar B \to X_s \gamma) $ by Heavy Flavor Average Group is \cite{HFAG}
\begin{equation}
 \mathcal B^{\rm exp}(\bar B \to X_s \gamma)=(3.55 \pm 0.24 \pm 0.09) \times 10 ^{-4}.
  \label{bav}
   \end{equation}
On  the theoretical side, the NLO calculation has been completed \cite{Misiak,Buras}, and gives
\begin{equation}
\mathcal B(\bar B \to X_s \gamma) = (3.57 \pm 0.30) \times 10 ^{-4}.
\end{equation}
The recent estimation at NNLO \cite{Misiak1} gives
$\mathcal B(\bar B \to X_s \gamma)=(3.15 \pm 0.23) \times 10 ^{-4}$,
which is about $1\sigma$ lower than the experimental average in Eq.~(\ref{bav}). Thus the experimental measurement of $\mathcal B(\bar B \to X_s \gamma)$ is in good  agreement with the SM predictions with roughly $10 \%$ errors on each side. The agreement would  provide strong constraints on the top quark anomalous interactions beyond the SM \cite{wtb,Fox}.

The decay amplitude of $t \to c \gamma$ has been calculated up to NLO \cite{Li}. For a consistent treatment of the constraints from $t \to c \gamma$ and $b \to s \gamma$ decays, we use the NLO formulas in Ref.~\cite{Misiak} to calculate $\mathcal B ( \bar B \to X_s \gamma )$. The experimental inputs and main formulas are collected in Appendix B.
For numerical analysis, we will use the notation
$ \kappa _{\rm tcR} ^{\gamma}=|\kappa _{\rm tcR} ^{\gamma}|e^{i \theta _{\rm tcR} ^{\gamma}}$ and set $\Lambda=1$ TeV. 

\begin{figure}
\centering
\includegraphics [width=7.5cm]{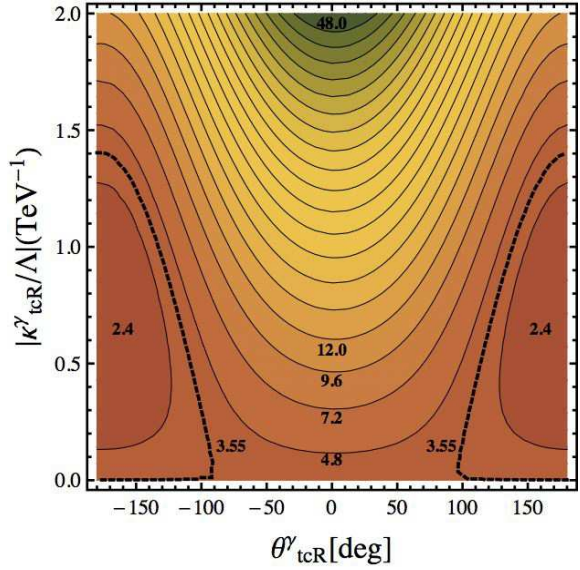}
\caption{The contour-plot  describes the dependence of
 $\mathcal B (\bar B \to X_s \gamma)(\times 10^{-4})$ on $ | \kappa _{\rm{tcR}} ^{\gamma} /\Lambda |$
 and $\theta _{\rm{tcR}} ^{\gamma}$. The dashed lines correspond to the experimental center
  value of $\mathcal B (\bar B \to X_s \gamma)$. }\label{fig2}
\end{figure}

At first, we analyze the dependence of $\mathcal B^{\rm SM+NP}(\bar B \to X_s \gamma)$ on the new physics parameters
$|\kappa _{\rm{tcR}} ^{\gamma} / \Lambda |$ and $\theta _{\rm{tcR}} ^{\gamma}$, which is shown in Fig.~\ref{fig2}.
From the figure,  one can find that   the contribution of anomalous
$t\to c \gamma$ coupling is constructive to the SM one  for $\theta _{\rm{tcR}} ^{\gamma} \in [-50^{\circ}, 50^{\circ}]$,
thus $\mathcal B (\bar B \to X_s \gamma)$ is very sensitive to $|\kappa _{\rm{tcR}} ^{\gamma}|$.
 However, when $|\theta _{\rm{tcR}} ^{\gamma}|\in [80^{\circ}, 130^{\circ}]$,
  the sensitivity  of $\mathcal B (\bar B \to X_s \gamma)$ to $|\kappa_{\rm{tcR}} ^{\gamma}|$ becomes weak.
    For   $|\theta _{\rm{tcR}}^{\gamma}|\sim  180^{\circ}$,  the contribution of anomalous $t\to c \gamma$ coupling is destructive to the SM one and there are two separated  possible strengths for $|\kappa _{\rm{tcR}} ^{\gamma} / \Lambda |$.
\begin{figure}[!ht]
\centering
\includegraphics[width=7.5cm]{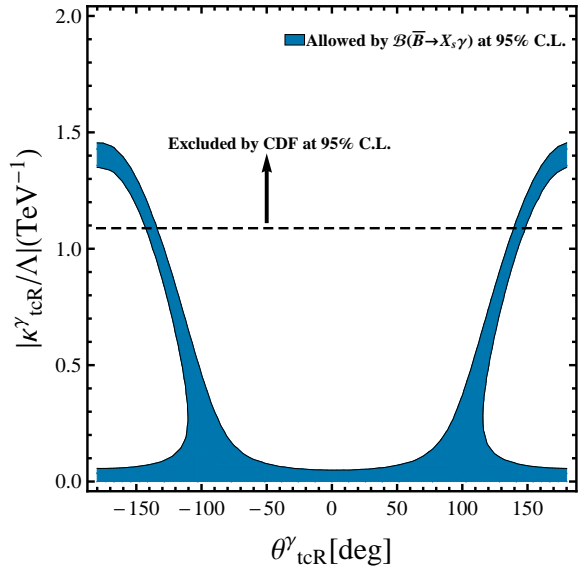}
\caption{The $95\%$ C.L. upper bounds on anomalous coupling $|\kappa _{\rm{tcR}} ^{\gamma} / \Lambda|$  as a function of $\theta_{\rm{tcR}} ^{\gamma}$.
 The shadowed region is allowed by $\mathcal B^{\rm exp}(\bar B \to X_s \gamma)$ and  the dash-line is the CDF \cite{CDF} upper limit.}
 \label{fig3}
\end{figure}
\begin{figure}[!ht]
\centering
\includegraphics [width=7.5cm]{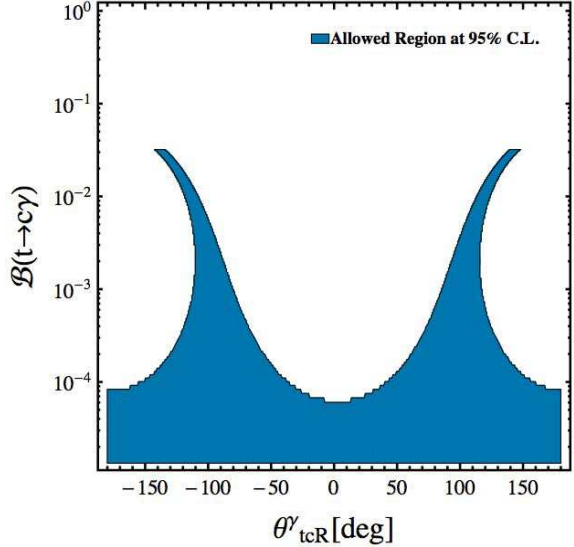}
\caption{  $\mathcal B(t \to c \gamma)$ as a function of $\theta _{\rm{tcR}}^{\gamma}$. The shadowed region is allowed by the combined constraints of $\mathcal B(\bar B \to X_s \gamma)$
and CDF searching at 95\% C.L.}
\label{fig4}
\end{figure}

The allowed region for the parameters $|\kappa _{\rm{tcR}}^{\gamma}/ \Lambda|$ and $\theta _{\rm{tcR}}^{\gamma}$ under the constraints from $\mathcal B(\bar B \to X_s \gamma)$ at $95\%$ C.L. is shown in Fig.~\ref{fig3}. The corresponding $95\%$ C.L. upper bound on  $\mathcal B(t \to c \gamma)$ is shown in Fig.~\ref{fig4}.

Now we turn to discuss the our numerical results.  From  Eq. (\ref{c7np}), the explicit relation  
 between the SM and the $t\to c \gamma$ coupling  contributions  is
 \begin{equation}
C_{7\gamma}^{\prime \rm{eff}}(m_{b})=-0.293 - 0.726~ \kappa_{\rm{tcR}}^{\gamma}.
\label{ka}
\end{equation}
Obviously,  when  $\rm{Re}~ \kappa _{\rm{tcR}}^{\gamma}>0$, the interference between them is constructive, 
and it turns to be destructive   when $\theta _{\rm{tcR}}^{\gamma} > 90^{\circ}$.   Thus the features of these constraints shown in Figs. \ref{fig3} and \ref{fig4} for different 
$\theta _{\rm{tcR}}^{\gamma}$ are 

\begin{enumerate}
\item[(i)]  the  bound on $|\kappa _{\rm{tcR}}^{\gamma}/ \Lambda|$ is  very strong for 
 $\theta _{\rm{tcR}}^{\gamma}\in [-50^{\circ}, 50^{\circ}] $.  For $\theta _{\rm{tcR}} ^{\gamma} \approx  0^{\circ}$,  as shown in Fig.~\ref{fig3}, we obtain the most restrictive  upper bound  $|\kappa _{\rm{tcR}} ^{\gamma}/ \Lambda| < 4.9 \times 10 ^{-5} \ \rm{GeV}^{-1}$,  which  implies   $\mathcal B(t \to c \gamma) < 6.54 \times 10^{-5}$; 

 \item[(ii)]   the  bound on $|\kappa _{\rm{tcR}}^{\gamma}/ \Lambda|$ is rather weak for $\theta _{\rm{tcR}}^{\gamma}$ around 
$110^{\circ}$.  For such a case, $\rm{Re}~\kappa_{\rm{tcR}}^{\gamma} )$ is destructive to the SM contribution as shown by 
Eq. (\ref{ka}), so,  the allowed strength for the  anomalous coupling is much larger than the one  for real $\kappa _{\rm{tcR}} ^{\gamma}$.   When  $|\theta _{\rm{tcR}}^{\gamma}| \approx 135^{\circ}$ and $|\kappa _{\rm{tcR}}^{\gamma}|\approx 0.571$,  $C_{7\gamma}^{\prime \rm{eff}}(m_{b})$  is almost  imaginary since 
${\rm Re} ~C_{7\gamma}^{\prime \rm{eff}}(m_{b}) \approx 0$.  
   Then the  restriction on $|\kappa_{\rm{tcR}} ^{\gamma} / \Lambda|$ is  provided by the CDF search for $\mathcal B(t \to c \gamma)$ \cite{CDF};
 
\item[(iii)]  as shown in Fig.~\ref{fig3},  when $\theta _{\rm{tcR}}^{\gamma} \sim \pm 180^{\circ}$, there are two solutions for $|\kappa _{\rm{tcR}} ^{\gamma}/ \Lambda|$. 
The larger one  $| \kappa^{\gamma}_{\rm{tcR}}/\Lambda | \sim 1.4 \times 10^{-3} ~\rm{GeV}^{-1}$(S2 column in Table \ref{table}) corresponds to the situation that the sign of $C^{\rm{eff}}_{7\gamma}$  is flipped.  However, it has been excluded by
the CDF upper bound of $\mathcal B( t \to c \gamma) <0.032$ \cite{CDF}.
The  another solution (S1 column in Table~\ref{table})  $|\kappa _{\rm{tcR}} ^{\gamma}/ \Lambda| < 5.6 \times 10^{-5} \ \rm{GeV}^{-1}$  will result in the upper limit $\mathcal B(t \to c \gamma) < 8.52 \times 10^{-5}$.
  \end{enumerate}

 Taking $\theta^{\gamma}_{\rm{tcR}}=0^{\circ},~\pm 180^{\circ}$ and $\pm 110^{\circ}$ as benchmarks, we summarize
our numerical constraints on $\kappa^{\gamma}_{\rm{tcR}}$ and their corresponding upper limits on $\mathcal B (t\to c\gamma)$  in Table~\ref{table}. From the table, we can find that our indirect bound on real $\kappa^{\gamma}_{\rm{tcR}}$ is much stronger  than the CDF direct bound.  The corresponding  upper limits on  $\mathcal B(t \to c \gamma)$ are  about the same  order as the ATLAS sensitivity   $\mathcal B(t \to c \gamma) > 9.4\times 10 ^{-5}$ \cite{ATLAS} and  CMS sensitivity
$\mathcal B(t \to c \gamma) > 4.1\times 10 ^{-4}$ \cite{CMS} with an integrated luminosity of $10 \ \rm{fb}^{-1}$ of the LHC operating at  ${\sqrt s} =14$ TeV \cite{ATLAS}.
%
\begin{table}[t]
\caption{  The 95\% C.L. constraints  on the  anomalous $t\to c\gamma$ coupling by  $\mathcal B(\bar B \to X_s \gamma)$ and $\mathcal B(t \to c \gamma)$ for some specific  $\theta_{\rm{tcR}}^{\gamma}$ values. }
\begin{center}
  \begin{tabular}{  l  c  c  c c c c c}
    \hline\hline
    & $\theta_{\rm{tcR}} ^{\gamma} = 0^{\circ}$ & $\theta_{\rm{tcR}} ^{\gamma} = \pm 180^{\circ} $ S1 & $ \theta_{\rm{tcR}} ^{\gamma} = \pm 180^{\circ}$ S2 & $\theta_{\rm{tcR}}^{\gamma}= \pm 110^{\circ}$    \\ \hline
    $\mathcal B(\bar B \to X_s \gamma)$ & $|\kappa_{\rm{tcR}}^{\gamma}| < 0.049$ & $|\kappa_{\rm{tcR}}^{\gamma}| < 0.056$ & $1.35 < |\kappa_{\rm{tcR}}^{\gamma}|<1.45$ &$|\kappa_{\rm{tcR}}^{\gamma}| < 0.55$  \\
    $\mathcal B(t \to c \gamma)$ \rm CDF  bounds\cite{CDF}&$|\kappa_{\rm{tcR}} ^{\gamma}| < 1.09
    \enskip $ & $|\kappa_{\rm{tcR}} ^{\gamma}| < 1.09\enskip$ & $|\kappa_{\rm{tcR}} ^{\gamma}| < 1.09$ & $|\kappa_{\rm{tcR}} ^{\gamma}| < 1.09$   \\
    Combined bounds   & $|\kappa_{\rm{tcR}}^{\gamma}| < 0.049$ & $|\kappa_{\rm{tcR}}^{\gamma}| < 0.056$ & $ -       $ & $| \kappa_{\rm{tcR}} ^{\gamma}| < 0.55$ \\
    $\mathcal B(t \to c \gamma) $  & $< 6.54 \times 10^{-5}$ & $< 8.52 \times 10^{-5}$ & $ -       $ & $<8.17 \times 10 ^{-3}$  \\ \hline \hline
  \end{tabular}
\end{center}
\vskip -0.5cm
\label{table}
\end{table}
\section{Conclusions}
In this paper, starting with model independent  dimension five  anomalous $t c \gamma$ operators, we have studied
their  contributions to $\mathcal B (\bar B \to X_{s} \gamma)$.  It is noted that
 the $t\to c \gamma$ transition will involve two independent operators
 $ \kappa ^\gamma _{\rm{tcR}} \bar c_L \sigma^{\mu \nu} t_R F_{\mu \nu}$ 
 and  $\kappa ^\gamma _{\rm{tcL}} \bar c_R \sigma^{\mu \nu} t_L F_{\mu \nu}$. The first operator will produce a left-handed photon in  $t \to c \gamma$ decay, while the second one will produce a right-handed photon.
It is found that $\bar B \to X_s \gamma$ is  sensitive to  the first operator, but not to the second  one.   

For real $\kappa ^{\gamma} _{\rm{tcR}}$, the constraint on the presence of   
$ \kappa ^\gamma _{\rm{tcR}} \bar c_L \sigma^{\mu \nu} t_R F_{\mu \nu}$ 
 is very  strong, which corresponds to the indirect upper limits
 $\mathcal B (t \to c\gamma)<6.54 \times 10^{-5}$ (for positive $\kappa ^{\gamma} _{\rm{tcR}}$) and
 $\mathcal B (t \to c\gamma)<8.52 \times 10^{-5}$ (for negative  $\kappa ^{\gamma} _{\rm{tcR}}$), respectively.
 These upper limits for  $\mathcal B (t \to c\gamma)$ are close to the $5 \sigma$ discovery  sensitivities  of ATLAS \cite{ATLAS} and  slightly lower than that of  CMS \cite{CMS}  with $10 \ \rm{fb}^{-1}$ integrated luminosity operating 
 at $\sqrt s =14$ TeV.  
 For  nearly  imaginary $\kappa ^\gamma _{\rm{tcR}}$,  the constraints are rather weak since $C_{7\gamma}$ in the SM is 
 a real number.  If $\mathcal B (t \to c\gamma)$ were found to be of  the order of 
  $\mathcal O (10^{-3})$ at the LHC  in the future, it would  imply the weak phase of  $\kappa ^\gamma _{\rm{tcR}}$ 
  to be around $\pm 100^{\circ}$. However, such a coupling might be ruled out by the other observable in B meson decays \cite{xqli}. 
  
  In summary, we have studied the interesting interplay between the precise measurement of $b\to s \gamma$ decay at B factories  and  the possible $t\to c \gamma$ decay at the LHC. For real anomalous coupling, it is shown that  $\mathcal B (t \to c\gamma)$ has been restricted to be blow $10^{-4}$ at $95\%$ C.L.  by  $\bar B \to X_{s}\gamma$ decay, which is already two order lower than the direct upper bound from CDF \cite{CDF}.   The result also implies that one may need data sample much larger than   $10 \ \rm{fb}^{-1}$ to hunt out $t\to c \gamma$ signals at the LHC.  
  
\section*{ACKNOWLEDGMENTS}
The work is supported by National Natural Science Foundation under contract
Nos.11075059 and 10735080. We thank Xinqiang Li for many helpful discussions and cross-checking calculations.
\newpage
\appendix
\begin{figure}[ht]
\centering
\subfigure[]{\includegraphics [width=4cm]{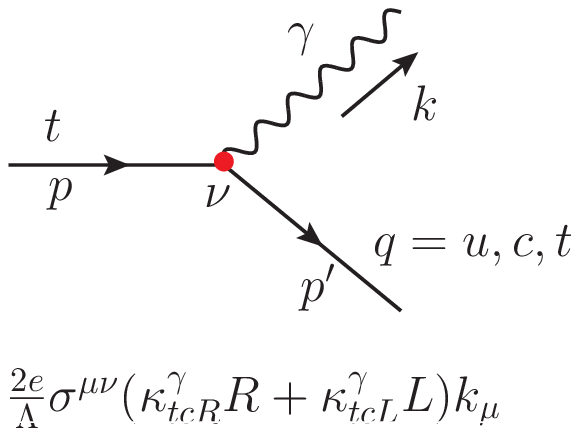}\label{fig5a}}
\subfigure[]{\includegraphics [width=5cm]{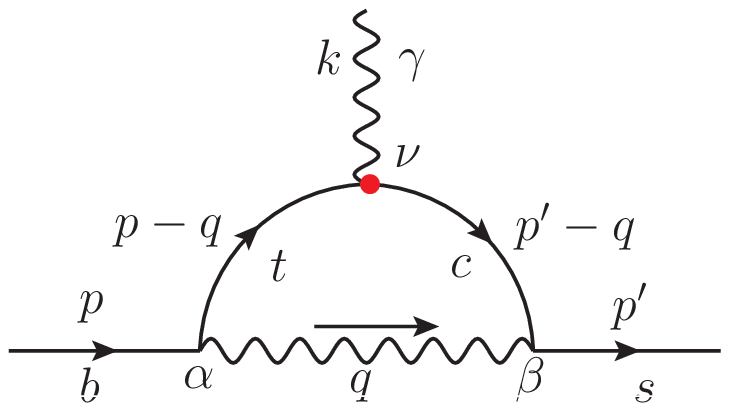}\label{fig5b}}
\caption{(a) the Feynman rules of $t \gamma c$ interactions in the Lagrangian of Eq.~\ref{lag1}. (b) penguin diagram contribution to $b \to s \gamma$ with top quark anomalous interactions.}\label{fig6}
\end{figure}
\section{The calculation of $C_{7 \gamma} ^{\rm NP}(\mu_{W})$}
\setcounter{equation}{0}
Using the Feynman rules in Fig~\ref{fig5a}, the amplitude of penguin diagram in Fig~\ref{fig5b} can be written as,
\begin{align}
i \mathcal{M} &= \bar u _s(p')[e \Gamma ^{\nu} (p,k)]u_b (p) \epsilon _{\nu}(k),
\\
\Gamma ^{\nu} (p,k) &= -  \frac {i g^2}{\Lambda} V ^{*} _{cs} V_{tb} \int \frac {d^4 q}{(2 \pi)^4} \frac {N}{[(p\prime -q)^2 -m _c ^2 + i \epsilon][(p-q)^2 - m_t ^2 +i \epsilon][q^2 - m _W ^2 + i \epsilon]},
\\
N&=\gamma _{\alpha} L (\dsl p\prime- \dsl q +m_q) \sigma^{\mu \nu} (\kappa^\gamma _{\rm{tcR}}R+\kappa^\gamma _{\rm{tcL}}L)(\dsl p -\dsl q +m_t)\gamma_{\beta} L g^{\alpha \beta} k_{\mu},
\end{align}
with $R=(1+\gamma^5)/2$ and $L=(1-\gamma^5)/2$. By Dirac algebra
\begin{eqnarray}
\gamma_{\alpha} L \dsl q \sigma ^{\mu \nu} (\kappa^\gamma _{\rm{tcR}}R+\kappa^\gamma _{\rm{tcL}}L)\dsl q\gamma_{\beta}L = 0,
\end{eqnarray}
the terms with $q^2$ in $N$ vanishes and N becomes
\begin{align}
N &= m_c \kappa^\gamma _{\rm{tcL}}[2(\dsl p - \dsl q) \sigma ^{\mu \nu} + (4-D)\sigma ^{\mu \nu}(\dsl p -\dsl q)]L k_{\mu} \nonumber \\
& \; +m_t \kappa^\gamma _{\rm{tcR}}[2 \sigma ^{\mu \nu}(\dsl p\prime -\dsl q) + (4 -D) (\dsl p \prime -\dsl q)\sigma ^{\mu \nu}]R k_{\mu}.
\end{align}
Thus, there is no divergence in $\Gamma ^{\nu}(p,k)$. After integrating out $q$ in the $\Gamma ^{\nu}(p,k)$ and using on-shell condition, $\Gamma ^{\nu} (p,k)$ can be written in the following form,
\begin{eqnarray}
e \Gamma^{\nu}(p,k)  = i e \frac {G_F}{4\sqrt 2 \pi^2} V^{*}_{cs} V_{tb}
\left[ i \sigma ^{\nu \mu} k_{\mu} (m_s f_{\rm{L}}(x) L + m_b f_{\rm{R}}(x) R)
\right],
\label{tcl}
\end{eqnarray}
where
  \begin{align}
    f_{\rm{L}}(x) = \frac{\kappa^{\gamma}_{\rm{tcL}}}{\Lambda} 2m_c
    \left[
    -\frac{1}{(x_c-1)(x_t-1)} - \frac{x_c^2}{(x_c-1)^2(x_c-x_t)} \ln x_c + \frac{x_t ^2}{(x_t -1)^2 (x_c -x_t)} \ln x_t
    \right],
    \label{a7}\\
    f_{\rm{R}}(x) = \frac{\kappa^{\gamma}_{\rm{tcR}}}{\Lambda} 2m_t
    \left[
    -\frac{1}{(x_c-1)(x_t-1)} - \frac{x_c^2}{(x_c-1)^2(x_c-x_t)} \ln x_c + \frac{x_t ^2}{(x_t -1)^2 (x_c -x_t)} \ln x_t\right],
  \end{align}
Using  the convention of Ref.~\cite{Buras}, we have
  \begin{align}
    C_{7 \gamma} ^{(0) \rm NP}(M_W) &= -\frac{1}{2} \frac{V _{cs} ^*}{V_{ts} ^*} f_{\rm{R}}(x)
    \nonumber \\
    &= \frac{V_{cs}^{*}}{V_{ts}^{*}} m_t \frac{\kappa ^\gamma _{\rm{tcR}}}{{\Lambda}} \left[\frac{1}{(x_c-1)(x_t-1)} + \frac{x_c ^2}{(x_c -1)^2(x_c - x_t)}\ln x_c - \frac{x_t ^2}{(x_t -1)^2(x_c -x_t)}\ln x_t \right].
  \end{align}
\section{ Main formulas and  inputs}
\setcounter{equation}{0}
Following the notation in Ref.~\cite{Misiak}, the branching ratio of $\bar B \to X_s \gamma$ can be expressed as
\begin{equation}
\mathcal B [\bar{B} \to X_s \gamma]_{E_{\gamma} > E_0}
= \mathcal B ^{\rm exp}[\bar{B} \to X_c e \bar{\nu}]
\left| \frac{ V^*_{ts} V_{tb}}{V_{cb}} \right|^2
\frac{6 \alpha_{\rm em}}{\pi\;C}
\left[ P(E_0) + N(E_0) \right],
\end{equation}
where $P(E_0)$ is the perturbative ratio
\begin{equation}
\frac{\Gamma[ \bar B \to X_s \gamma]_{E_{\gamma} > E_0}}{
|V_{cb}/V_{ub}|^2 \; \Gamma[ \bar B \to X_u e \bar{\nu}]} =
\left| \frac{ V^*_{ts} V_{tb}}{V_{cb}} \right|^2
\frac{6 \alpha_{\rm em}}{\pi} P(E_0),
\end{equation}
which includes the Wilson coefficients of Eq.~\ref{C7eff}. $N(E_0)$ denotes the non-perturbative corrections. The semileptonic phase space factor
\begin{equation}
C = \left| \frac{V_{ub}}{V_{cb}} \right|^2
\frac{\Gamma[\bar{B} \to X_c e \bar{\nu}]}{\Gamma[\bar{B} \to X_u e \bar{\nu}]}
\end{equation}
can be obtained from a fit of the experimental spectrum of the $\bar B \to X_c l \bar{\nu}$ \cite{C}.

For calculating $\mathcal B (t \to c \gamma)$, we use the NLO formulas in Ref.~\cite{Li} and \cite{Li1}. Because $t \to b W$ is the dominant top quark decay mode, the branching ratio of $t \to c \gamma$ is defined as
\begin{equation}
\mathcal B(t \to c \gamma) = \frac{\Gamma(t \to c \gamma)}{\Gamma(t \to bW)}.
\end{equation}
The partial width $\Gamma( t \to c \gamma)$ at the NLO can be found in Ref.~\cite{Li}, namely,
\begin{equation}
\Gamma_{\rm{NLO}}(t \to c \gamma)=\frac{2\alpha_s}{9\pi}\Gamma_0(t \to c\gamma)\left[-3\log\left(\frac{\mu^2}{m_t^2}\right)-2\pi^2+8\right],
\end{equation}
where $\Gamma_0(t \to c \gamma)= \alpha m_t^3\left(\kappa^{\gamma}_{\rm{tcR}} / \Lambda\right)^2$ is the LO partial decay width.

The partial width of $t \to b W$ has been calculated in Ref.~\cite{Li1} at the NLO, which reads
\begin{align}
\Gamma_{\rm{NLO}}(t \to bW)=\Gamma_0(t \to bW) \biggl\lbrace 1+\frac{2\alpha_s}{3\pi}\biggl[2\left(\frac{(1-\beta_W^2)(2\beta_W^2-1)(\beta_W^2-2)}
{\beta_W^4(3-2\beta_W^2)}\right)\ln(1-\beta_W^2)  \nonumber \\
-\frac{9-4\beta_W^2}{3-2\beta_W^2}\ln\beta_W^2 +2\mathrm{Li}_2(\beta_W^2) -2\mathrm{Li}_2(1-\beta_W^2)-\frac{6\beta_W^4-3\beta_W^2-8}{2\beta_W^2(3-2\beta_W^2)}-\pi^2 \biggr]\biggr\rbrace
\end{align}
with $\Gamma_0(t \to bW) = \frac{G_Fm_t^3}{8\sqrt{2}\pi}|V_{tb}|^2\beta_W^4(3-2\beta_W^2)$ and $\beta_W\equiv(1-m_W^2/m_t^2)^{1/2}$.
\begin{table}
\caption{Experimental inputs for calculating the branching ratio of $\bar B \to X_s \gamma$ and $t \to c \gamma$.}
\begin{center}
  \begin{tabular}{c c}
    \hline \hline
    \multicolumn{2}{c}{Experimental Inputs} \\
    \hline
    $\alpha_{em}=1/137.036$            \cite{PDG} & $M_Z = 91.1876 \pm 0.0021 \ \rm GeV$  \cite{PDG} \\
    $\alpha_s(M_Z)=0.1184 \pm 0.0007$  \cite{PDG} & $M_W = 80.399 \pm 0.023 \ \rm GeV$    \cite{PDG} \\
    $G_{\rm F} = 1.16637 \times 10^{-5} \ \rm GeV ^{-2}$  \cite{PDG} & $m_b^{\rm 1S} = 4.67_{-0.06}^{+0.18} \ \rm GeV$  \cite{PDG}\\
    $A = 0.812 ^{+0.013} _ {-0.027}         $  \cite{CKMfitter}& $m_c(m_c) = (1.224 \pm 0.017 \pm 0.054) \ \rm GeV$ \cite{Manohar}\\
    $\lambda = 0.22543 \pm 0.00077          $  \cite{CKMfitter}& $m_{t,pole} = 172.0 \pm 0.9 \pm 1.3 \ \rm GeV$  \cite{PDG}\\
    $\bar {\rho} = 0.144 \pm 0.025 $ \cite{CKMfitter}& $\mathcal B ^{\rm exp}[\bar{B} \to X_c e \bar{\nu}] = ( 10.64 \pm 0.17 \pm 0.06 )\%$ \cite{BABAR}\\
    $\bar {\eta} = 0.342 ^{+0.016} _{-0.015}$  \cite{CKMfitter}&$C=0.580 \pm 0.016$ \cite{C}\\
    $\left|V^*_{ts} V_{tb} / V_{cb} \right|^2 = 0.9625$ & $\epsilon_{\rm ew} = 0.0071 $  \cite{Misiak,Gambino}\\
    $(V_{us}^{*}V_{ub}) / (V_{ts}^{*}V_{tb}) = -0.007 + 0.018 \rm i$ & $N(E_0) = 0.0036 \pm 0.0006$  \cite{Misiak}\\
    $V _{cs} ^* / V_{ts} ^* = -24.023 - 0.432 \rm i$ &$E_0 = 1.6 \ \rm GeV$ \\
    \hline \hline
  \end{tabular}
\end{center}
\label{table2}
\end{table}

The experimental inputs are collected in Table.~\ref{table2}, in which the CKM factors are derived from the Wolfenstein parameters A, $\lambda$, $\bar \rho$ and $\bar \eta$.

\end{document}